\documentclass[a4paper,fleqn]{cas-dc}

\usepackage[authoryear]{natbib}

\usepackage[switch]{lineno}

\begin{document}
\let\WriteBookmarks\relax
\def\floatpagepagefraction{1}
\def\textpagefraction{.001}
\shorttitle{HiPERCAM characterisation of 2022 OB$_5$}
\shortauthors{M.R. Alarcon et al.}

\title [mode = title]{Accessible does not mean exploitable: HiPERCAM reveals the ultra-fast rotation of 2022 OB$_5$}                      



\author[1,2,3]{M. R. Alarcon}[orcid=0000-0002-8134-2592]
\cormark[1]
\ead{mra@mralarcon.com}
\credit{Conceptualization, Methodology, Data analysis, Discussion, Writing -- original draft, Writing -- review \& editing}

\author[2,3]{J. Licandro}[orcid=0000-0002-9214-337X]
\credit{Discussion, Writing -- review \& editing}

\author[1,2,3]{M. Serra-Ricart}[orcid=0000-0002-2394-0711]
\credit{Discussion, Writing -- review \& editing}

\author[2,4]{D. Garcia-Álvarez}[orcid=0009-0007-3475-3376]
\credit{Observation, Data preparation, Writing -- review \& editing}

\author[2,4]{A. Cabrera-Lavers}[orcid=0000-0002-9153-8724]
\credit{Observation, Writing -- review \& editing}

\affiliation[1]{
                organization={Light Bridges},
                addressline={Observatorio Astronómico del Teide}, 
                city={Güímar},
                postcode={38570}, 
                state={Tenerife}, 
                country={Spain}
                }

\affiliation[2]{
                organization={Instituto de Astrofísica de Canarias (IAC)},
                addressline={C/ Vía Láctea, s/n}, 
                city={La Laguna},
                postcode={38205}, 
                state={Tenerife}, 
                country={Spain}
                }

\affiliation[3]{
                organization={Departamento de Astrofísica, Universidad de La Laguna (ULL)},
                addressline={C/ Vía Láctea, s/n}, 
                city={La Laguna},
                postcode={38206}, 
                state={Tenerife}, 
                country={Spain}
                }
\affiliation[4]{
                organization={GRANTECAN},
                addressline={Cuesta de San José s/n}, 
                city={Breña Baja},
                postcode={38712}, 
                state={La Palma}, 
                country={Spain}
                }

\cortext[cor1]{Corresponding author}

\begin{abstract}
2022 OB$_5$ is a sub-10-metre Apollo-type near-Earth asteroid whose orbital configuration placed it among the most dynamically accessible small bodies in near-Earth space, motivating its selection as the target of the first commercial asteroid-prospecting mission. We present its first photometric characterisation, based on high-cadence simultaneous five-band $u_sg_sr_si_sz_s$ observations obtained with HiPERCAM at the 10.4-m Gran Telescopio Canarias (GTC). Analysis of the light curves yields a rotation period of $P_{\rm rot} = 1.542 \pm 0.001$ min, independently confirmed with observations taken by the Two-meter Twin Telescope, establishing 2022 OB$_5$ as an ultra-fast rotator. The reflectance spectrum derived from the simultaneous multiband photometry is featureless and moderately red, consistent with the X-complex. Despite its good orbital accessibility, the ultra-fast rotation of 2022 OB$_5$ poses severe practical challenges for any surface operation with current technology, regardless of compositional interest. This illustrates a population-level challenge: at the sizes and $\Delta v$ values most favourable for in-situ missions, fast rotation is the dominant spin state, and rotation period measurement is therefore an indispensable prerequisite for evaluating the resource potential of asteroid mission candidates.
\end{abstract}



\received{26 March 2026}
\revised{23 April 2026}
\accepted{4 May 2026}


\maketitle

\section{Introduction}

Near-Earth asteroids (NEAs) span a wide range of sizes, compositions, and internal structures, yet their physical properties remain poorly constrained across most of the size spectrum, lagging far behind the rapidly growing rate of new discoveries. This mismatch is most pronounced at the smallest end of the known population: metre- to decametre-sized objects are numerous, dynamically accessible, and thought to be closely connected to the source populations of meteorites \citep{Binzel2004, Harris2015, Granvik2018, DeMeo2022}, but they are still very poorly characterised owing to their faintness, rapid motion, and typically brief visibility windows from ground-based observatories.

In addition to their scientific interest, some NEAs have also attracted attention as potential targets for in-situ resource utilisation, since their proximity to Earth and unusually low transfer requirements make them natural candidates for future resource-prospecting missions \citep{Barbee2010, Elvis2011}. A small subset can be reached from low Earth orbit with a comparatively low velocity increment $\Delta v$, making them among the most dynamically accessible small bodies in near-Earth space. In this context, metal-rich objects are of particular interest. They may host substantial Fe--Ni metal and elevated concentrations of platinum-group elements, materials that are both geochemically scarce and economically valuable on Earth \citep{Kargel1994, Schubert2025}. Iron meteorites, widely regarded as fragments of differentiated asteroid cores, demonstrate that such enrichments are physically achievable and can, in some cases, exceed the abundances found in terrestrial ore deposits \citep{Cannon2023}. Identifying metal-rich bodies from remote observations nonetheless remains a significant challenge, as the spectral signatures of metallic surfaces are largely featureless at visible wavelengths and easily confused with those of other compositional classes \citep{Tholen1984, Harris2014}.

Among these objects, 2022 OB$_5$ stands out as a particularly compelling case. It is a sub-10-metre Apollo-type NEA discovered on 2022 July 29 by the MAPS survey in San Pedro de Atacama, Chile. With an absolute magnitude of $H=29$ and a minimum orbit intersection distance (MOID) with Earth of only 0.0041~au, its orbital configuration makes it a dynamically favourable target: the estimated mission $\Delta v$ is $\sim 3.6$ km s$^{-1}$, well within the range commonly adopted to define mission-accessible NEAs \citep{Thirouin2016}. These orbital properties attracted attention beyond the planetary science community. In early 2025, the private company AstroForge announced 2022 OB$_5$ as the target of its Odin mission, the first-ever commercial asteroid-prospecting mission \citep{Astroforge}. The Odin spacecraft was subsequently lost during its deep-space cruise phase; a follow-up mission is currently in preparation. The mission was conceived as a prospecting flyby motivated by the possibility that the object could be metal-rich. At the time of our observations, no rotation period, light-curve amplitude, or spectroscopic data had been published for 2022 OB$_5$. Its rotational state, a fundamental prerequisite for assessing the viability of any surface operation, therefore remained unknown.

Obtaining those constraints is, however, observationally challenging. Small NEAs commonly exhibit rotation periods of minutes or even seconds \citep{Beniyama2022, Devogele2024}, so colour measurements obtained through sequential, non-simultaneous filter observations can be significantly biased by rotational variations, potentially affecting taxonomic classifications derived from broadband photometry \citep{Alarcon2026}. High-cadence observations are therefore essential for accurately sampling the light curves of potential fast rotators. At the same time, simultaneous multi-band imaging provides the most robust way to measure colours without rotational bias and thus to recover the intrinsic reflectance properties of the target. Instruments capable of acquiring several photometric bands in a single exposure are therefore especially valuable for the taxonomic characterisation of rapidly rotating NEAs \citep{Licandro2023}.

In this work we present a photometric characterisation of 2022 OB$_5$ based on observations obtained with HiPERCAM \citep{Dhillon2021}, a quintuple-beam high-speed imager mounted on the 10.4-m Gran Telescopio Canarias (GTC). HiPERCAM records simultaneous $u_sg_sr_si_sz_s$ photometry with negligible dead time and second-level cadence, making it especially well suited to the study of faint, rapidly rotating, and fast-moving Solar System targets. Its combination of large-aperture sensitivity, simultaneity across five optical bands, and sub-second temporal sampling enables the derivation of both a robust rotational light curve and instantaneous colours free from the ambiguities inherent to sequential observations.

This work represents the first application of HiPERCAM to asteroid characterisation. These data provide the first direct observational constraints on 2022 OB$_5$, an object already selected as a candidate target for commercial deep-space prospecting, and illustrate the broader potential of high-cadence simultaneous multiband photometry for the characterisation of the smallest NEAs. The observations and reduction methods are described in Section~\ref{sec:met}. The derived rotation period, light-curve properties, and visible reflectance spectrum are presented in Section~\ref{sec:results} and discussed in Section~\ref{sec:discussion} in the context of the small-NEA population and mission-accessible targets.

\section{Observations and Methods}
\label{sec:met}

\subsection{HiPERCAM} \label{sec:met/HiPERCAM}
The near-Earth asteroid 2022 OB$_5$ was observed on 2026 January 14 using HiPERCAM \citep{Dhillon2021}, a quintuple-beam, high-speed optical imager mounted at the Folded Cass-G focus of the 10.4-m Gran Telescopio Canarias (GTC), located at the El Roque de Los Muchachos Observatory (ORM; La Palma, Spain). HiPERCAM achieves simultaneous five-band ($u_s g_s r_s i_s z_s$) imaging through a series of four dichroic beamsplitters that direct light toward five custom-designed, back-illuminated Teledyne e2v CCD288 sensors. These $2048 \times 1024$ pixel frame-transfer devices are specifically engineered to provide near-zero dead time between exposures: the readout time between consecutive frames is 8~ms, resulting in a duty cycle of $\sim 99.2$\% for the 1~s exposures used in this run. As no mechanical shutter is employed, the image transfer to the storage area is near-instantaneous and introduces no differential illumination pattern across the detector.

The observations started on 2026 January 14 at 22:57 UTC and ended at 23:40 UTC, spanning 42 minutes. At that time, 2022 OB$_5$ was at a geocentric distance of 0.00434~au (1.69 lunar distances) and at a solar phase angle of 42 deg. The asteroid exhibited an apparent visual magnitude of 18.8 and a significant apparent motion of 38 arcsec min$^{-1}$. The observations were conducted in non-sidereal tracking mode following the specific orbital solution of the target. To sample the asteroid's light curve while avoiding trailing and maximizing cadence, the exposure time was 1 s, which ensured that both the asteroid and field stars remained point-like throughout the 42-minute run. The CCDs were operated in Slow readout mode to minimize read-out noise while maintaining high-speed capability and the data were acquired with $2 \times 2$ pixel binning. This configuration resulted in an effective pixel scale of 0.16 arcsec pixel$^{-1}$ over a field of view of approximately $2.8 \times 1.4$ arcmin. The observations were performed under clear sky conditions with an average airmass of 1.11 and a stable seeing around 1.0 arcsec.

\subsection{Instrumental flux extraction} \label{sec:met/phot_sources}
Standard calibration frames were obtained at the end of the observing night to ensure instrumental consistency. Master bias frames were constructed for each CCD by combining 101 zero-exposure frames using a $2.5\sigma$-clipped median algorithm. Twilight sky flats, originally acquired with $1 \times 1$ binning, were spatially resampled to match the $2 \times 2$ science configuration. These frames were bias-subtracted and individually normalized by their median flux before being combined into a final master flat per band using a $3\sigma$-clipping median filter. All science images were subsequently bias-subtracted and divided by their corresponding master flats.

The detection and centroiding of point sources were performed frame-by-frame across the entire field of view. To ensure a reliable identification of sources in bands with lower signal-to-noise ratio (S/N), particularly in the $u_s$ filter, the detection process was executed on images previously convolved with a Gaussian kernel. This spatial filtering suppressed high-frequency noise, allowing for stable centroiding of both the asteroid and all detectable field stars despite the short exposure times. Once the coordinates were determined on the filtered frames, a global photometric extraction was performed on the original calibrated images to avoid noise correlations and preserve the native statistical properties of the data.

Fixed-aperture photometry was performed with a radius of 9 pixels, corresponding to $1.5 \times$ the average FWHM. The local sky background was estimated using a circular annulus with inner and outer radii of $4\times$ and $6\times$ FWHM, respectively, where a $3\sigma$-clipped median was applied to ensure robustness against faint stellar contamination. The total variance for each measurement was modeled as:
\begin{equation} 
\sigma_{F_{e^-}}^{2} =
F_{e^-}
+ n_{\mathrm{ap}}\sigma_{\mathrm{sky},e^-}^{2}
+ \frac{n_{\mathrm{ap}}^{2}}{n_{\mathrm{sky}}}\sigma_{\mathrm{sky},e^-}^{2}, \label{eq:var}
\end{equation}
where $F_{e^-}$ is the background-subtracted source flux in electrons, $n_{\mathrm{ap}}$ is the number of pixels within the source aperture, $n_{\mathrm{sky}}$ is the number of pixels used to estimate the local sky background and $\sigma_{\mathrm{sky},e^-}$ is the standard deviation of the sky counts per pixel in the annulus, measured in electrons. These terms represent, respectively, the source Poisson noise, the background noise within the aperture and the uncertainty in the determination of the local sky level. This is exact for a mean sky estimator and represents an approximation when a $3\sigma$-clipped median is used instead, as is the case here; the difference is negligible in practice for the sky backgrounds encountered in our data.

\subsection{Ensemble relative photometry} \label{sec:met/phot_rel}
After the photometric catalogues for each band were produced, point sources detected in consecutive frames were linked into tracks using a nearest-neighbour matching scheme. Because the observations were performed in non-sidereal tracking mode, the telescope followed the apparent motion of the asteroid and the field stars drifted across the detector at the same apparent rate as the target ($\sim$38 arcsec min$^{-1}$). To account for this global image motion, a frame-to-frame translation $(\delta x_j,\delta y_j)$ was estimated before each matching step by computing the median positional offset of all mutually matched sources between frames $j-1$ and $j$. The predicted position of each active track in frame $j$ was then obtained by displacing its last measured position by this global offset and a final association radius of 12 pixels was used to confirm individual links. Detections that could not be associated with any active track were registered as new tracks. Because field stars enter and leave the field of view during the 42-minute run, tracks of varying length are expected.

A per-frame photometric correction term $Z_j$ was derived from an ensemble of stable field stars in order to remove frame-to-frame
transparency variations. Only tracks spanning at least $L_{\min}$ frames and exhibiting an internal scatter, estimated via the median absolute deviation of their instrumental magnitudes, below a band-dependent threshold $\sigma_{\max}$ were retained as photometric reference stars. Given the lower S/N in the $u_s$ band, relaxed thresholds of $L_{\min} = 5$ and $\sigma_{\max} = 0.15$ mag were adopted for that band, while values of $L_{\min} = 8$ and $\sigma_{\max} = 0.10$ mag were used for the remaining bands. These selections yielded 293, 305, 492, 649 and 656 reference tracks in the $u_s$, $g_s$, $r_s$, $i_s$ and $z_s$ bands, respectively, reflecting the increasing number of detectable field stars toward redder wavelengths.

For each frame $j$, the zero-point contribution of reference star $i$ was computed as
\begin{equation}\label{eq:zp_contrib}
    z_{ij} = m_{ij} - \tilde{m}_i,
\end{equation}
where $m_{ij} = -2.5\log_{10}(F_{e^-,ij})$ is the instrumental magnitude of star $i$ in frame $j$ and $\tilde{m}_i$ is
the median instrumental magnitude of that star over the entire track. The frame zero-point $Z_j$ was then estimated as the
$3\sigma$-clipped median of $\{z_{ij}\}$ over all reference stars contributing to frame $j$, with a formal uncertainty of
\begin{equation}
    \sigma_{Z_j} = \frac{1.2533\;\hat{\sigma}_{Z_j}}{\sqrt{n_j}},
    \label{eq:zp_err}
\end{equation}
where $\hat{\sigma}_{Z_j}$ is the normalized median absolute deviation (NMAD) of the clipped residuals and $n_j$ is the number of stars used. Frames with fewer than $n_{\min}$ contributing stars (3 for $u_s$, 5 for the remaining bands) were flagged as unmeasured and their zero-point values were linearly interpolated from the nearest valid frames. The fraction of frames with a directly measured zero-point was 62\% in $u_s$, 91\% in $g_s$ and above 99\% in $r_s$, $i_s$ and $z_s$.

The corrected instrumental magnitude of the asteroid in frame $j$ relative to the stars ensemble is
\begin{equation}
    m_{\mathrm{rel},j} = m_{\mathrm{inst},j} - Z_j,
    \label{eq:mrel}
\end{equation}
with a total photometric uncertainty
\begin{equation}
    \sigma_{m,j} = \sqrt{\sigma_{\mathrm{f},j}^{2} + \sigma_{Z_j}^{2}},
    \label{eq:merr}
\end{equation}
where $\sigma_{\mathrm{f},j}$ is the photometric uncertainty propagated into magnitude units from the variance model of Eq. \ref{eq:var}.

\subsection{Period search and outlier rejection} \label{sec:met/period}
An estimate of the rotation period of 2022 OB$_5$ is obtained to phase-fold the light curves and identify and reject outlying data points prior to the scientific analysis. The period search was carried out on the $r_s$ differential light curve, which provides the largest number of valid data points and the highest signal-to-noise ratio among the five HiPERCAM bands. An initial period estimate was obtained using the generalised Lomb--Scargle periodogram \citep{Lomb1976,Scargle1982}, using a multi-term Fourier model with $K = 4$ harmonics to capture the non-sinusoidal shape.

Them Lomb--Scargle solution was subsequently refined using Phase Dispersion Minimization \citep[PDM;][]{Stellingwerf1978}. Once the best period was established, the light curve was phase-folded using
\begin{equation}
    \phi_i = \left(\frac{t_i - t_0}{P_{\rm rot}}\right) \bmod 1,
    \label{eq:phase}
\end{equation}
where $t_0 = 61054.9568023$~MJD is the time of the first valid observation and $P_{\rm rot}$ is the refined PDM period. A second outlier-rejection step was applied to the folded reference light curve by flagging points deviating by more than $3\sigma$ from the median within each of 50 equal phase bins. The cleaned phase-folded light curve was then fitted with a fourth-order Fourier series,
\begin{equation}
    m(\phi) = a_0 + \sum_{k=1}^{4}
              \left[a_k \cos(2\pi k\phi) + b_k \sin(2\pi k\phi)\right],
    \label{eq:fourier}
\end{equation}
which provides a smooth template of the rotational modulation. This template was subsequently used to analyse the light curves obtained in the remaining four HiPERCAM bands.

For each band, the differential light curve $m_{\mathrm{rel}}$ was phase-folded using the template and the reference epoch $t_0$. A scalar offset $C_b$ was then determined by iteratively minimising the median residual
\begin{equation}
    \mathrm{Res}_{i,b} = m_{\mathrm{rel},i} - \left[m(\phi_i) + C_b\right],
    \label{eq:offset}
\end{equation}
after which residual outliers were identified through $3\sigma$ clipping within the same 50 phase bins. The reference model provides an excellent description of the phase-folded light curve in all five bands after the offset correction, with around 6\% of the points rejected as outliers. The magnitude offsets $C_b$ contain contributions from both the inter-band photometric zero-point differences and the intrinsic colour of the asteroid.

\subsection{Absolute photometric calibration} \label{sec:met/calib}
\begin{figure*}
    \centering
    \includegraphics[width=\textwidth]{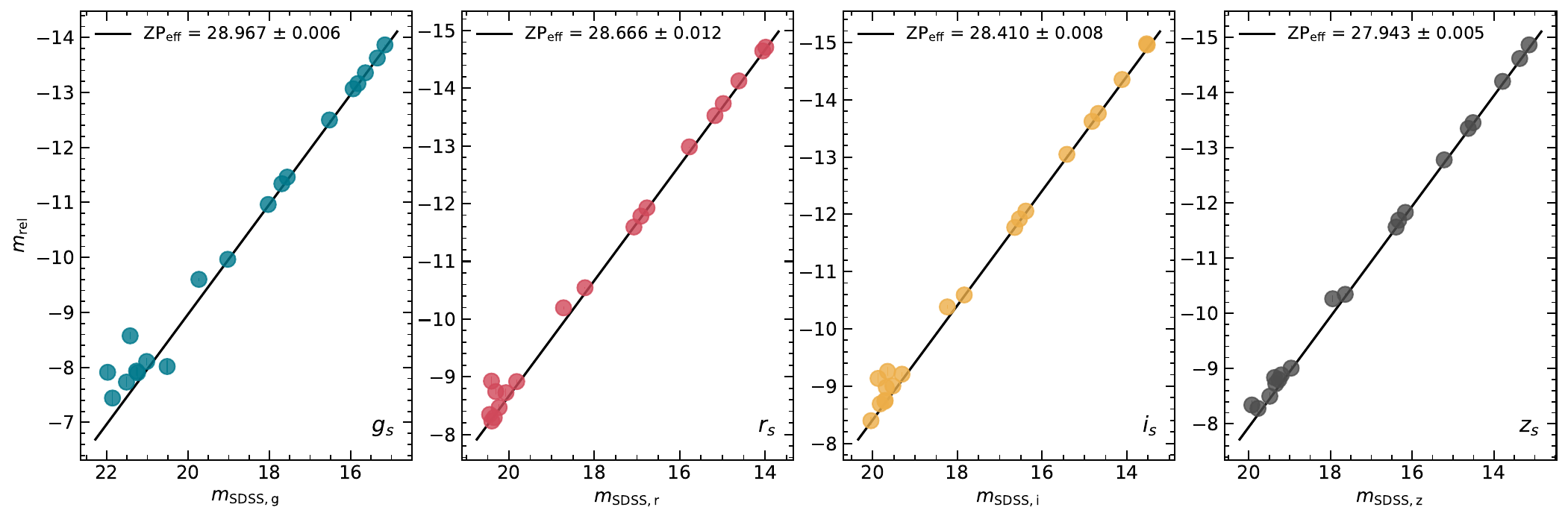}
    \caption{Corrected instrumental magnitude $m_{\rm rel,\star}$ versus SDSS magnitude for field stars in the reference frame, for the $g_s$, $r_s$, $i_s$ and $z_s$ HiPERCAM bands. The solid line in each panel shows the best-fit calibration relation, with the derived effective zero-point $ZP_{\rm eff}$ indicated.}
    \label{fig:zpcal}
\end{figure*}

Absolute photometric calibration was performed independently for each band using field stars detected in a single reference frame. The pixel coordinates of a set of isolated, unsaturated field stars were identified in each CCD and their equatorial coordinates determined by fitting a six-parameter affine transformation with their corresponding positions from the Gaia DR3 catalogue \citep{Gaia2023}, achieving astrometric residuals below $0.11^{\prime\prime}$.

The field stars were cross-matched against photometric catalogues to derive absolute zero-points. Ideally, this would be performed against the Sloan Digital Sky Survey \citep[SDSS;][]{SDSS2012}, whose $ugriz$ filter system most closely matches that of HiPERCAM. However, the observed field lies outside the SDSS footprint. We therefore used the Pan-STARRS DR2 catalogue \citep[PS1;][]{PS12016}, which provides uniform all-sky coverage in $grizy$ to comparable depth  
and a photometric uncertainty below 0.05 mag. Then, the PS1 magnitudes were converted to the SDSS system using the colour-dependent transformations of \citet{Finkbeiner2016}:
\begin{equation}
    m_{\rm PS1} - m_{\rm SDSS} = \sum_{k=0}^{3} a_k x^k,
    \qquad x \equiv (g-i)_{\rm PS1},
    \label{eq:finkbeiner}
\end{equation}
where the coefficients $a_k$ are tabulated in \citet{Finkbeiner2016} and the transformation is valid for main-sequence stars with $0.4 < (g-i)_{\rm PS1} < 2.7$. 

The differential light curves of the asteroid are referenced to the median transparency level of the entire run via the ensemble zero-point $Z_j$. To ensure a consistent photometric reference, the instrumental magnitudes of the field stars
in the reference frame were corrected to the same nightly median level before computing the zero-point:
\begin{equation}
    m_{\rm rel,\star} = m_{\rm inst,\star}  - Z_{\rm ref},
    \label{eq:mrel_star}
\end{equation}
The effective zero-point for each band was then estimated as the $3\sigma$-clipped mean of the differences between the SDSS magnitudes and the corrected instrumental magnitudes,
\begin{equation}
    ZP_{\rm eff,b} = \overline{\left(m_{\rm SDSS,b} -
    m_{\rm rel,\star,b}\right)},
    \label{eq:zpeff}
\end{equation}

Figure \ref{fig:zpcal} shows, for each of the $g_s r_s i_s z_s$ bands, the relative magnitude of the field stars as a function of their SDSS magnitude, together with the best-fit calibration relation. The residuals have an rms below 0.02 mag in $r_s$ and $i_s$, rising to $\sim$0.05 mag in $g_s$ and $z_s$, consistent with the larger colour-term corrections required in those bands. As an independent validation, the calibrated magnitudes of a separate set of field stars in a different frame of the observing run were compared to their SDSS magnitudes. We found residuals of $\lesssim$0.04 mag in all four bands, with a typical scatter of $\sim$0.02 mag, confirming that the calibration is accurate and stable throughout the observing run.

No reliable photometric catalogue with $u$-band coverage is available for the field of view: the SDSS footprint does not encompass the observed region and Pan-STARRS does not observe in this filter. The colour transformation from g$_{\rm PS1}$ to u$_{\rm SDSS}$ in \citet{Finkbeiner2016} is strongly sensitive to stellar metallicity and was therefore not used for calibration. Instead, the effective zero-point in $u_s$ was derived from the nightly zero-point provided by the GTC facility,
\begin{equation}
    ZP_{{\rm eff},u_s} = ZP_{{\rm tel},u_s} - k_u X_{\rm ref} + Z_{{\rm ref},u_s},
    \label{eq:zpeff_u}
\end{equation}
where $ZP_{{\rm tel},u_s} = 27.96 \pm 0.01$ mag is the zero-point at airmass zero, $k_u = 0.38 \pm 0.05$ mag airmass$^{-1}$ is the atmospheric extinction coefficient for the ORM observatory \citep{King1985}, $X_{\rm ref}$ is the airmass of the reference frame and the term $Z_{{\rm ref},u_s}$ brings the facility zero-point to the same nightly reference level as the other bands.

\section{Results} \label{sec:results}

\subsection{Rotation period}\label{sec:res/rotation}

\begin{figure}
    \centering
    \includegraphics[width=\columnwidth]{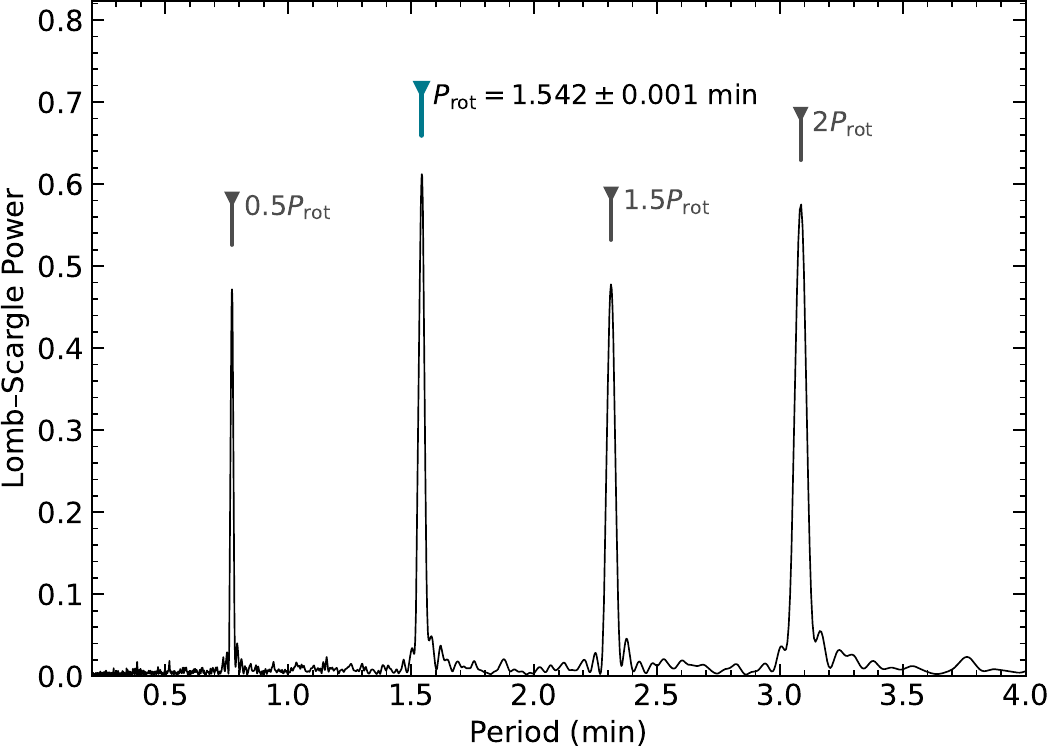}
    \caption{Generalised Lomb--Scargle periodogram of the $r_s$ differential light curve of 2022 OB$_5$. The dominant peak at $P_{\rm rot} = 1.542 \pm 0.001$ min (blue marker) corresponds to the best-fit rotation period. The remaining peaks at $0.5 P_{\rm rot}$, $1.5 P_{\rm rot}$, and $2 P_{\rm rot}$ are aliases arising from the non-sinusoidal shape of the light curve in combination with the multi-term Fourier model.}
    \label{fig:periodogram}
\end{figure}

\begin{figure*}
    \centering
    \includegraphics[width=.8\textwidth]{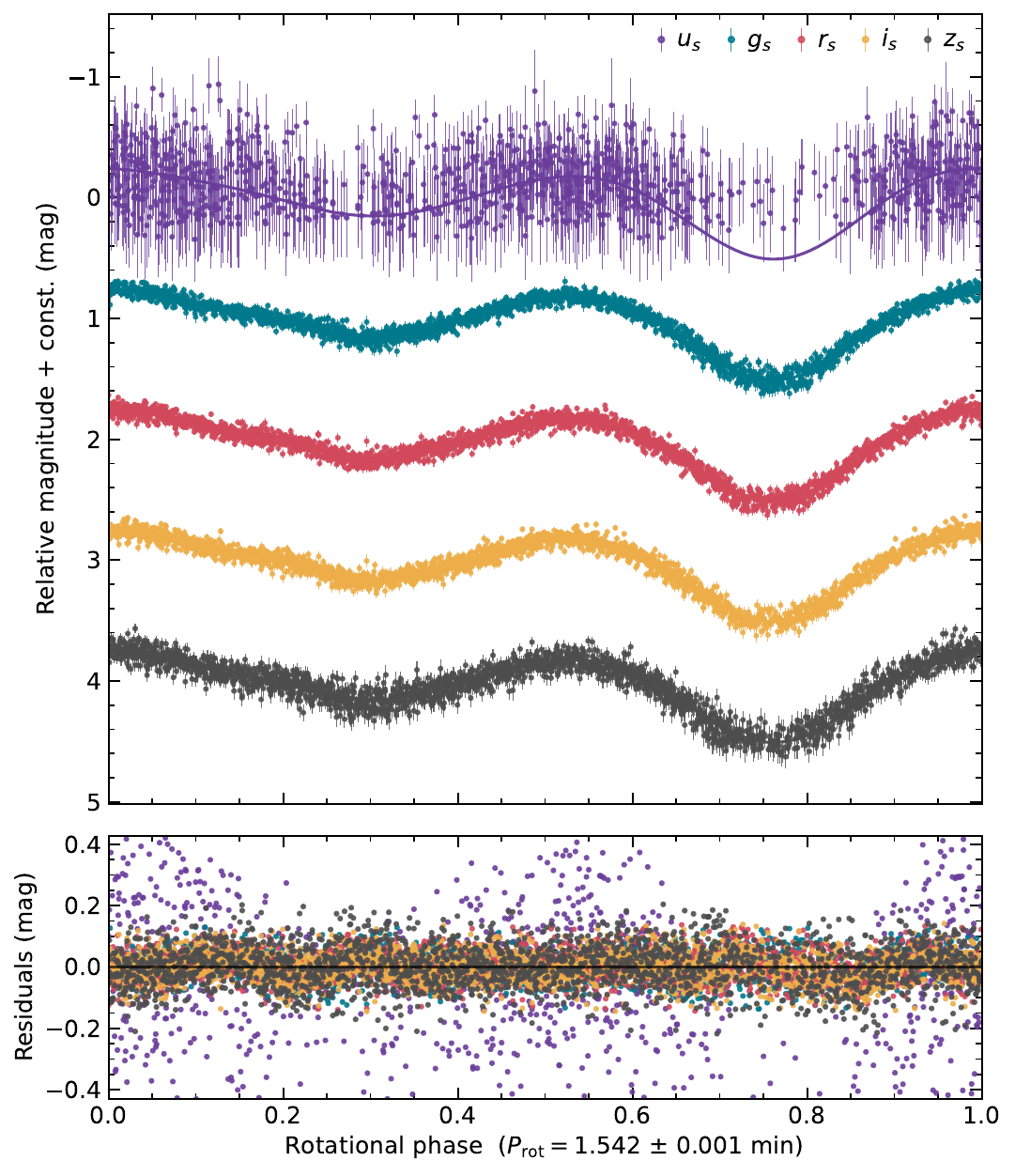}
    \caption{Phase-folded light curves of 2022 OB$_5$ obtained simultaneously in the five HiPERCAM bands ($u_s$, $g_s$, $r_s$, $i_s$, $z_s$), displayed from top to bottom in order of increasing wavelength and offset vertically for clarity. The solid lines show the best-fit fourth-order Fourier model derived from the $r_s$ band and applied to the remaining bands. The sparse coverage near the primary minimum in the $u_s$ band reflects the low signal-to-noise ratio of this filter, where the flux of the target cannot be reliably measured at the faintest phases. The bottom panel shows the residuals with respect to the model for all five bands. The rotation period is $P_{\rm rot} = 1.542 \pm 0.001$ min. The phase is computed using the reference epoch is $t_0 = 61054.9568023$~MJD.}
    \label{fig:lc}
\end{figure*}

The rotation period of 2022 OB$_5$ was determined from the $r_s$ differential light curve following the procedure described in Section \ref{sec:met/period}. The generalised Lomb--Scargle periodogram, combined with PDM, yields a best-fit rotation period of $P_{\rm rot} =  1.542 \pm 0.001$ min. The periodogram is shown in Figure~\ref{fig:periodogram}, where the dominant peak is unambiguous and the  remaining peaks correspond to aliases arising from the non-sinusoidal shape of the light curve. This is strictly a synodic period. However, given the short observing window and the orbital configuration of 2022 OB$_5$ at the time of the observations, the difference between the synodic and sidereal periods is of order $10^{-6}$~min and is therefore negligible for the purposes of this analysis.

Figure \ref{fig:lc} shows the phase-folded light curves in all five HiPERCAM bands, displayed from top to bottom in order of increasing wavelength. The rotational modulation is clearly detected in all bands, with a peak-to-trough amplitude of $\Delta m \approx 0.75$ mag. Assuming a triaxial ellipsoidal shape viewed equator-on, the observed amplitude must first be corrected to zero phase angle following
\begin{equation}
    A(0) = \frac{A(\alpha)}{1 + m\cdot\alpha},
    \label{eq:amp_corr}
\end{equation}
where $\alpha = 42.4$ deg is the solar phase angle at the time of the observations and $m = 0.013$ mag deg$^{-1}$ is the slope parameter appropriate for M- and C-type asteroids \citep{Zappala1990}. This yields a corrected amplitude $A(0) = 0.48$ mag. The lower limit on the axis ratio is then
\begin{equation}
    \frac{a}{b} \geq 10^{0.4  A(0)},
    \label{eq:axisratio}
\end{equation}
yielding $a/b \geq 1.56$. No colour-dependent modulation is observed across the five bands, indicating that the light curve shape and amplitude are consistent regardless of the band. The light curve is well described by a single-period Fourier model with no evidence of secondary periodicities or irregular amplitude variations, suggesting that 2022 OB$_5$ is rotating in a principal-axis state with no detectable tumbling motion at frequencies high enough to be discernible within the 42-minute observing run.

\begin{figure}
    \centering
    \includegraphics[width=\columnwidth]{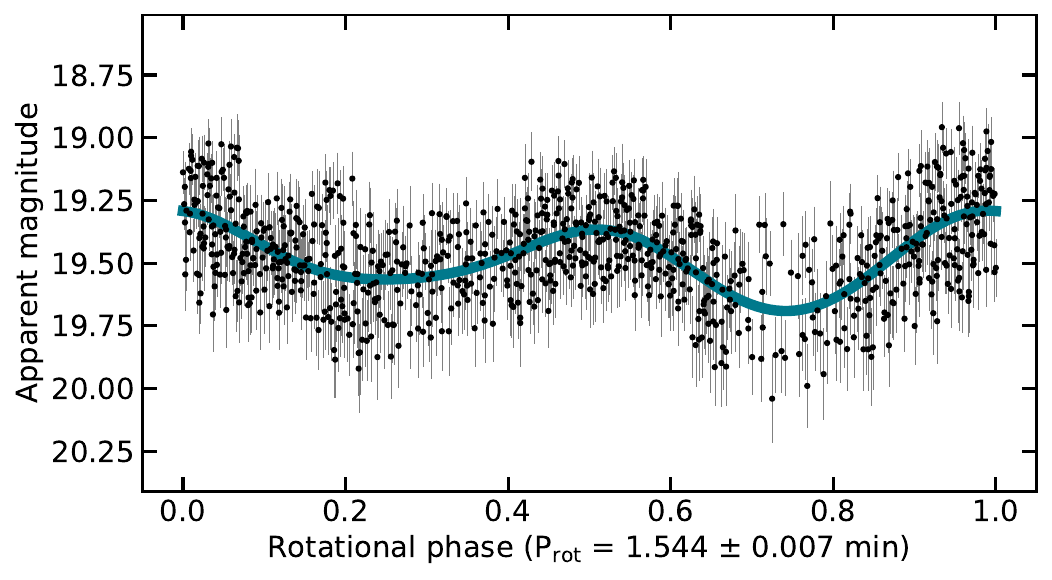}
    \caption{Phase-folded light curve of 2022 OB$_5$ obtained on 2026 January 14 with TTT3, one of the two 2.0-m telescopes of the Two-meter Twin Telescope (TTT) facility at the Teide Observatory. The solid line shows the best-fit Fourier model, yielding a rotation period of $P_{\rm rot} = 1.544 \pm 0.007$ min. The phase is computed using the reference epoch is $t_0 = 61054.9568023$~MJD.}
    \label{fig:lc_ttt3}
\end{figure}

As an independent validation, the target was also observed on the same night with TTT3, one of the two 2.0-m telescopes of the Two-meter Twin Telescope\footnote{\hyperlink{ttt.iac.es}{ttt.iac.es}} facility, located at the Teide Observatory (Tenerife, Spain) and equipped with the FERVOR instrument, an optical imager based on a sCMOS sensor IMX455 \citep{Alarcon2023}. Observations were performed through a broadband $g^\prime + r^\prime$ filter with an exposure time of 2 s. The phase-folded light curve is shown in Figure \ref{fig:lc_ttt3}. The rotation period derived from these data, $P_{\rm rot} = 1.544 \pm 0.007$ min, is in excellent agreement with the HiPERCAM solution, providing a robust confirmation of the period determination. The observational strategy and data reduction pipeline for TTT3 data are described in detail in \citet{Alarcon2026}.

2022 OB$_5$ falls firmly in the ultra-fast rotator regime, spinning well above the cohesionless spin barrier. The implications of this result for the internal structure and cohesion of 2022 OB$_5$ and its place within the population of small NEAs are discussed later.

\subsection{Colours and taxonomic classification}
\label{sec:res/taxonomy}
The simultaneous five-band photometry provided by HiPERCAM allows us to derive the spectral reflectance of 2022 OB$_5$ directly from its calibrated light curves. The colour of the asteroid in each band $b$ relative to $r$ was computed as
\begin{equation}
    (b - r)_{\rm ast} = (C_b - C_{r_s}) + (ZP_{\rm eff,b} - ZP_{\rm eff,r_s}),
    \label{eq:color}
\end{equation}
where $C_b$ is the magnitude offset defined in Eq. \ref{eq:offset} and $ZP_{\rm eff,b}$ is the effective photometric zero-point of Eq. \ref{eq:zpeff}. The first term captures the relative brightness of the asteroid between bands as measured from the ensemble-referenced light curves, while the second converts these differential offsets to the absolute SDSS photometric system. The derived colours are:
\begin{equation*}
\begin{aligned}
    u - r &= +1.98 \pm 0.06 \\
    g - r &= +0.525 \pm 0.013 \\
    i - r &= -0.157 \pm 0.014 \\
    z - r &= -0.193 \pm 0.013
\end{aligned}
\end{equation*}
The normalised reflectance spectrum was then computed as
\begin{equation}
    R_b = 10^{-0.4\left[(b-r)_{\rm ast} - (b-r)_{\odot}\right]},
    \label{eq:reflectance}
\end{equation}
where $(b - r)_{\odot}$ are the solar colours in the SDSS photometric system \citep{Holmberg2006}, so that $R_{r} = 1$ by construction.

\begin{figure}
    \centering
    \includegraphics[width=\columnwidth]{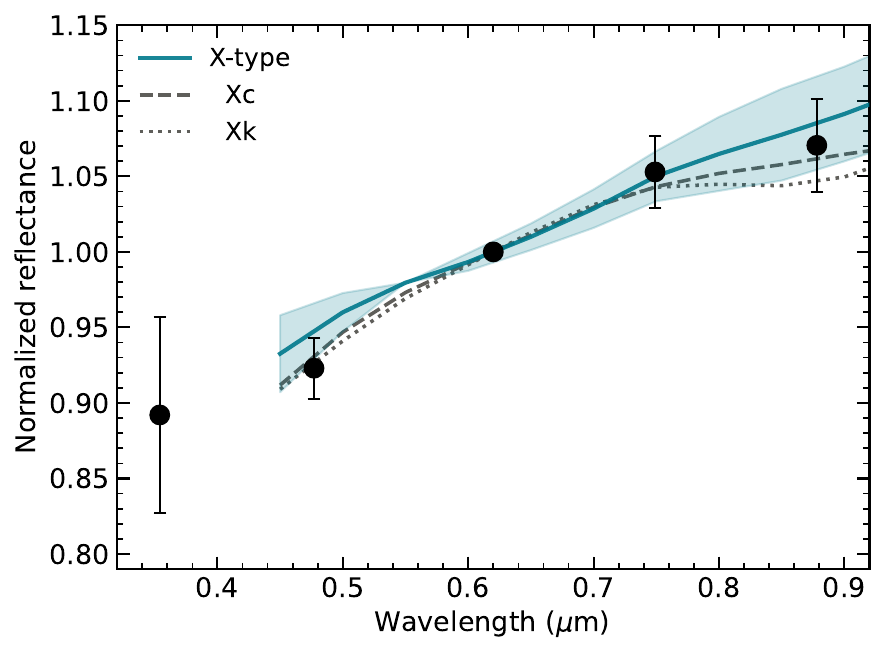}
    \caption{Normalised reflectance spectrum of 2022 OB$_5$ derived from simultaneous HiPERCAM photometry in the $u_s g_s r_s i_s z_s$ bands. The solid line and shaded region show the mean reflectance spectrum and its uncertainty for the X-type taxonomic class from the Bus-DeMeo classification system \citep{BusDeMeo2009}. The dashed and dotted lines show the Xc and Xk subtypes, respectively, which provide the best match to the observed reflectance spectrum.}
    \label{fig:refl}
\end{figure}

The resulting reflectance spectrum of 2022 OB$_5$ is shown in Figure \ref{fig:refl}. It is characterised by a slightly sub-solar slope in the blue optical and a gently rising continuum toward the red, yielding an overall featureless and moderately red spectral shape
across the wavelength range covered by HiPERCAM. Comparison with the Bus-DeMeo \citep{BusDeMeo2009} taxonomic templates shows the best overall agreement with the X-complex and among its subtypes, with the Xc and Xk classes.

\section{Discussion}
\label{sec:discussion}
\begin{figure*}
    \centering
    \includegraphics[width=0.49\textwidth]{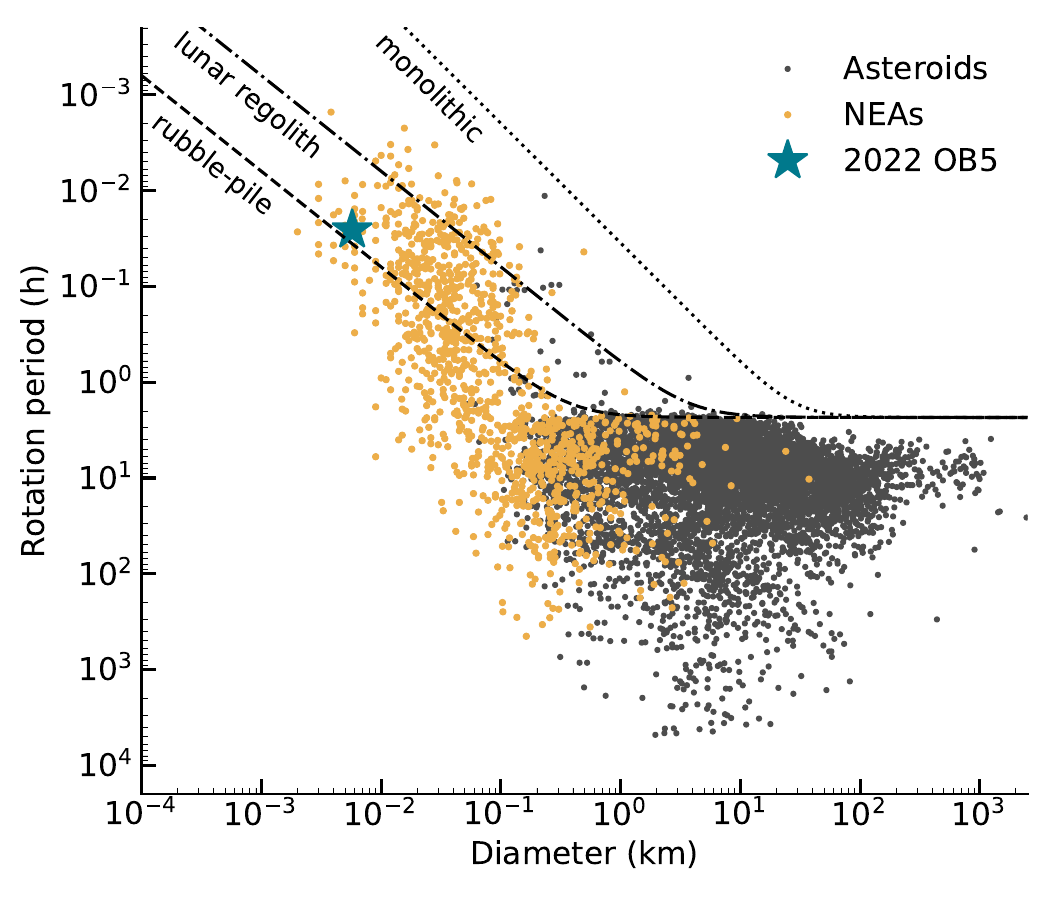}
    \hfill
    \includegraphics[width=0.49\textwidth]{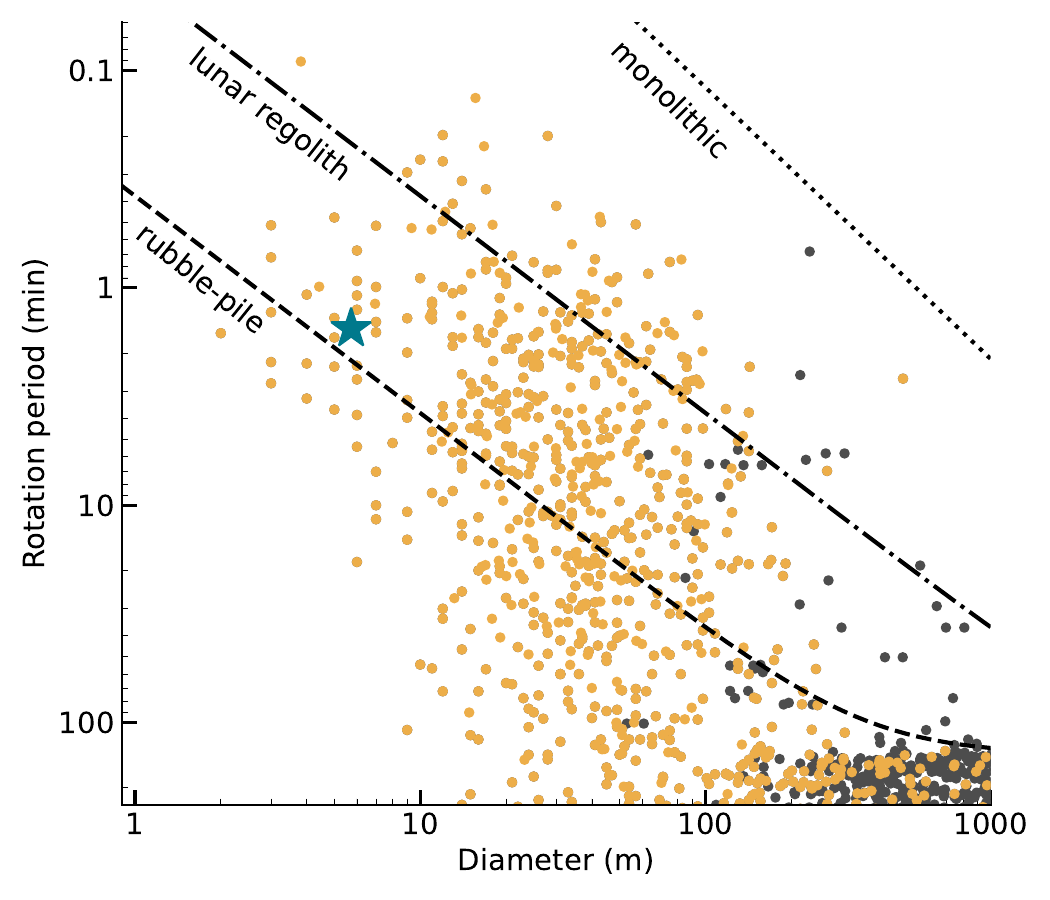}
    \caption{Rotation period as a function of diameter. The left panel shows the full distribution, with asteroids from the Lightcurve Database (LCDB) and SsODNet with $U \geq 2$ in grey and near-Earth asteroids (NEAs) from the same catalogues combined with the sample of \citet{Alarcon2026} in yellow. 2022 OB$_5$ is marked with a blue star. Diameters were estimated from absolute magnitudes assuming a geometric albedo $p_V = 0.15$. Cohesive spin limits for weak rubble-pile material ($C \sim 10$~Pa), lunar-like regolith ($C \sim 1$~kPa) and a size-dependent monolithic strength ($C \propto D^{-0.5}$) are overplotted for reference. The right panel provides a zoomed view of the sub-140~m size range.}
    \label{fig:spin_barrier}
\end{figure*}

The classification of 2022 OB$_5$ within the X-complex reflects the well-known compositional degeneracy of this group at optical wavelengths. In the Tholen taxonomy, the E, M and P classes occupy this same broad spectral domain and are distinguished primarily by albedo rather than by visible reflectance alone \citep{Tholen1984}. Within the Bus-DeMeo framework, our photometry is most consistent with the Xc subtype, characterised by a relatively low-to-moderate spectral slope and a mildly concave spectral curvature, while the next best match is Xk, which shares a similarly featureless continuum but may exhibit a shallow absorption band near 0.9 $\mu$m attributed to orthopyroxene \citep{BusDeMeo2009, Fornasier2011}. Photometric observations obtained by MANOS in February 2025 and reported recently classified the object as Xe \citep{Moskovitz2026}, a subtype commonly associated with high-albedo E-type asteroids and aubrite-like compositions. However, our broadband photometry does not support the presence of the characteristic $\sim$0.49~$\mu$m absorption feature of Xe objects, making this interpretation less favoured than Xc or Xk. A definitive subtype assignment would require near-infrared spectroscopy covering the $\sim$0.9 $\mu$m region, together with an independent albedo determination to break the E/M/P degeneracy. Accordingly, the bulk density remains poorly constrained and could plausibly span the wide range encompassed by X-complex analogues, ranging from $\rho \sim 1800$ kg m$^{-3}$ for low-albedo, porous P-type analogues to $\rho \sim 4900$ kg m$^{-3}$ for metal-rich M-type compositions \citep{Carry2012}.

This compositional ambiguity also has a direct impact on the size estimate of 2022 OB$_5$. The absolute magnitude reported by JPL Horizons is $H = 28.99$, though catalog values for small asteroids carry typical uncertainties of $\sim 0.5$ mag \citep{Pravec2012}. The albedo distributions of the three X-complex subtypes are markedly different: based on the photometric sample of \citet{DeMeoCarry2013}, E-type objects have $p_V = 0.536 \pm 0.247$ ($n = 47$), M-types $p_V = 0.143 \pm 0.051$ ($n = 825$), and P-types $p_V = 0.053 \pm 0.012$ ($n = 771$), though these values may be biased toward higher albedos due to the observational selection of the underlying sample. Combining the uncertainty in $H$ with the full range of plausible albedos, the effective diameter of 2022 OB$_5$ spans nearly one order of magnitude, from $\sim$2~m (for $p_V = 0.7$, $H = 29.5$) to $\sim$15~m (for $p_V = 0.03$, $H = 28.5$).

This X-type classification is consistent with the taxonomic trends observed at metre-class sizes. While the X-complex accounts for approximately $\sim$10\% of kilometre-scale NEAs \citep{Binzel2019}, its fraction rises systematically with decreasing size, reaching $\sim$17\% at $\lesssim$600 m \citep{Perna2018} and $\sim$24\% at $\lesssim$140 m \citep{Devogele2019}. This increase has been attributed to the growing contribution of Hungaria-family E-type objects at small sizes and to a discovery bias towards higher-albedo objects in this regime. The X-complex classification of 2022 OB$_5$ is therefore consistent with the observed enrichment of this group at metre-class sizes, while leaving the specific mineralogy open.

The density uncertainty inherent to the X-complex classification directly shapes the structural interpretation of the rotation period. At $P_{\rm rot} \sim 1.5$ min and $D \lesssim 10$ m, 2022 OB$_5$ lies between the weak rubble-pile and regolith-strength cohesion limits shown in Figure \ref{fig:spin_barrier}, which are derived from the Drucker--Prager yield criterion \citep{Holsapple2007, Zhang2021} and correspond to cohesive strengths of $C \sim 10$ Pa (van der Waals forces between micrometre-scale regolith grains; \citealt{Scheeres2010}) and $C \sim 1$ kPa (lunar-like regolith; \citealt{Sanchez2014}), respectively, computed for $\rho = 3$ g cm$^{-3}$ and a friction angle of $30^{\circ}$. Self-gravity alone is wholly insufficient at this spin rate for any plausible bulk density: the body must be held together by either cohesive inter-grain forces or material strength. If the composition tends towards the metallic end of the X-complex ($\rho \gtrsim 4000$ kg m$^{-3}$), the higher density raises the spin limits and relaxes the structural demands. At the intermediate densities typical of enstatite chondrite or mixed silicate compositions ($\rho \sim 2500$--$3600$ kg m$^{-3}$; \citealt{Carry2012}), van der Waals cohesion of the order of tens to hundreds of pascals, as expected for fine regolith grains \citep{Scheeres2010, Sanchez2014}, would be sufficient to maintain structural integrity. A monolithic or fractured-monolith interpretation is equally consistent with the data, as the observed spin rate lies well below the size-dependent monolithic strength limit across all plausible compositions. As with most objects in this size regime, the two scenarios cannot be disentangled from rotation rate alone without independent density or shape information.

The rotation period of 2022 OB$_5$ is not anomalous within the population of known ultra-fast rotators at comparable sizes. Dedicated surveys and individual characterisation campaigns have revealed that periods of one to a few minutes are commonplace in this size regime \citep{Beniyama2022, Licandro2023, Alarcon2026}, with the fastest known rotators reaching periods of just a few seconds. The $P_{\rm rot} = 1.542$~min measured here is therefore representative of the spin state expected for a sub-10-metre NEA, rather than an outlier, further supporting the view that fast rotation is the norm at these sizes.

Despite its exceptional orbital accessibility, the rotation period of 2022 OB$_5$ imposes a severe practical constraint on any surface operation. At $P_{\rm rot} \sim 1.5$ min, the centrifugal acceleration at the equator of an elongated body with $D \lesssim 10$ m with a bulk density of $\rho \sim 3$ g~cm$^{-3}$ is $a_c \approx 2 \times 10^{-2}$ m~s$^{-2}$, while the surface gravitational acceleration is $g \approx 4 \times 10^{-4}$ m~s$^{-2}$, giving a ratio $a_c/g \gtrsim 50$. The centrifugal acceleration thus exceeds surface gravity by nearly two orders of magnitude, making anchoring, sampling, or resource extraction highly challenging with current technology, regardless of the compositional interest. If 2022 OB$_5$ belongs to the metallic end of the X-complex, it could in principle be a source of iron-nickel or platinum-group elements, but the absence of albedo and near-infrared data prevents any quantitative evaluation of this possibility.

More fundamentally, the combination of an excellent $\Delta v$ and a physically inaccessible surface is not a peculiarity of this object: at the sizes and $\Delta v$ values most favourable for rendezvous missions, rapid rotation is the dominant spin state, with fast rotators comprising $\gtrsim 94$\% of objects with $H > 26$ \citep{Alarcon2026}. Only a small fraction of dynamically accessible targets simultaneously satisfy the rotation-period criterion $P > 1$ h considered viable for surface operations \citep{Thirouin2016}, a constraint that 2022 OB$_5$ fails by more than two orders of magnitude. 
The case of 2022 OB$_5$ illustrates a population-level challenge for asteroid resource missions: physical characterisation, and rotation period measurement in particular, is an indispensable step prior to any evaluation based on orbital accessibility or compositional criteria alone.

\section{Conclusions}

We have presented the first photometric characterisation of 2022 OB$_5$, a mission-accessible sub-10-metre Apollo-type NEA, based on simultaneous five-band $u_sg_sr_si_sz_s$ observations with HiPERCAM at the GTC, the first application of this instrument to a minor body in the Solar System.

The light curve analysis yields a rotation period of $P_{\rm rot} = 1.542 \pm 0.001$ min, independently confirmed by TTT3 observations, placing 2022 OB$_5$ firmly in the ultra-fast rotator regime. The simultaneous multiband photometry reveals a featureless, moderately red reflectance spectrum consistent with the X-complex, with Xc and Xk as the best-matching subtypes. A definitive compositional assignment requires near-infrared spectroscopy and an independent albedo determination.

Despite its orbital accessibility, the ultra-fast rotation of 2022 OB$_5$ poses severe practical constraints on any surface operation with current technology, regardless of its compositional interest. This is not a peculiarity of this object: rapid rotation dominates the spin distribution of dynamically accessible small NEAs, and rotation period measurement is therefore an indispensable step in the evaluation of any asteroid resource candidate.

\section*{Declaration of competing interest}
The authors declare that they have no known competing financial interests or personal relationships that could have appeared to influence the work reported in this paper.

\section*{Acknowledgments}
We thank the anonymous reviewers for their constructive comments, which helped improve the manuscript. This work is based on observations made with the Gran Telescopio Canarias (GTC), installed at the Spanish Observatorio del Roque de los Muchachos of the Instituto de Astrofísica de Canarias on the island of La Palma. It makes use of data obtained with the HiPERCAM instrument, built by the Universities of Sheffield, Warwick and Durham, the UK Astronomy Technology Centre, and the Instituto de Astrofísica de Canarias. Development of HiPERCAM was funded by the European Research Council, and its operations and enhancements by the Science and Technology Facilities Council. Additional observations were obtained with the Two-meter Twin Telescope (TTT), located at the Teide Observatory of the Instituto de Astrofísica de Canarias (IAC) and operated by Light Bridges in Tenerife, Canary Islands, Spain. The observation time rights (DTO) used for this research were consumed under the PEI “FASTROTA26”. This research also made use of storage and computing capacity at ASTRO POC’s EDGE computing centre in Tenerife in the form of Indefeasible Computer Rights (ICR), likewise consumed under the PEI “FASTROTA26”. Dr Antonio Maudes’s economic and legal insights were instrumental in shaping this work.
M.R.A. and J.L. acknowledge support from the Agencia Estatal de Investigaci\'on del Ministerio de Ciencia e Innovaci\'on (AEI-MCINN) under grant "Hydrated Minerals and Organic Compounds in Primitive Asteroids" with reference PID2020-120464GB-100.

\printcredits

\bibliographystyle{cas-model2-names}
\bibliography{cas-refs}

\end{document}